\documentclass[journal]{IEEEtran4PSCC}

\usepackage{textcomp}
\usepackage{tabu}
\usepackage{comment}
\usepackage{setspace}
\usepackage{gensymb}
\usepackage{cite}
\usepackage{amsmath,bm,amssymb}
\usepackage{mathtools}
\usepackage[utf8]{inputenc}
\usepackage{tikz}
\usepackage{array,arydshln}
\usepackage{lipsum}
\usepackage{epstopdf}
\usepackage{eurosym}
\usepackage{hhline}
\usepackage{epsfig}
\usepackage{caption}
\usepackage{mdwlist}
\usepackage{float}
\usepackage{subfig}
\usepackage{xurl}
\usepackage{algpseudocode}
\usepackage{algorithm}
\usepackage{url}
\usepackage{multirow}
\usepackage{comment}
\usepackage{color,soul}
\usepackage{hhline}
\usepackage{pgfplots}
\usepackage{graphicx,dblfloatfix}
\usepackage{circuitikz}
\usepackage{siunitx}
\usetikzlibrary{plotmarks}
\usetikzlibrary{patterns}

\newcolumntype{C}[1]{>{\centering\let\newline\\\arraybackslash\hspace{0pt}}m{#1}}

\makeatletter
\newcommand*\short[1]{\expandafter\@gobbletwo\number\numexpr#1\relax}
\makeatother

\usepackage[acronym, automake]{glossaries}
\makeglossaries
\newacronym{PV}{PV}{photovoltaic}
\newacronym{DER}{DER}{distributed energy resource}
\newacronym{DLNN}{DLNN}{deep learning neural network}
\newacronym{AMI}{AMI}{advanced metering infrastructure}
\newacronym{NN}{NN}{neural network}
\newacronym{BNN}{BNN}{Bayesian neural network}
\newacronym{DSSE}{DSSE}{distribution system state estimation}
\newacronym{WLS}{WLS}{weighted least squares}
\newacronym{DN}{DN}{distribution network}
\newacronym{DSO}{DSO}{distribution system operator}
\newacronym{LV}{LV}{low-voltage}
\newacronym{MV}{MV}{medium voltage}
\newacronym{EV}{EV}{electric vehicle}
\newacronym{SM}{SM}{smart meter}
\newacronym{QR}{QR}{quantile regression}
\newacronym{RMSE}{RMSE}{root mean square error}
\newacronym{SE}{SE}{state estimation}
\newacronym{PI}{PI}{prediction interval}
\newacronym{LVSE}{LVSE}{low-voltage state estimation}
\newacronym{FS}{FS}{feature set}
\newacronym{KL}{KL}{Kullback-Leibler}

\pgfplotsset{compat=newest}

\definecolor{color1}{RGB}{0,0,0} 
\definecolor{color2}{RGB}{182,193,251} 
\definecolor{bblue}{HTML}{4F81BD}
\definecolor{rred}{HTML}{C0504D}
\definecolor{ggreen}{HTML}{9BBB59}
\definecolor{ppurple}{HTML}{9F4C7C}
\definecolor{sandybrown}{HTML}{f4a460}
\definecolor{lightseagreen}{HTML}{20b2aa}
\definecolor{cornflowerblue}{HTML}{6495ed}
\definecolor{limegreen}{HTML}{32CD32}
\definecolor{orange}{HTML}{ffa500}
\definecolor{purple}{HTML}{6495ed}

\usepackage{pgfplots}
    \usetikzlibrary{
        pgfplots.dateplot,
    }
\usepgfplotslibrary{fillbetween}

\pgfplotsset{compat=1.11,
    /pgfplots/ybar legend/.style={
    /pgfplots/legend image code/.code={%
       \draw[##1,/tikz/.cd,yshift=-0.25em]
        (0cm,0cm) rectangle (3pt,0.8em);},
   },
}

\glsdisablehyper
\usepackage[font=footnotesize]{caption}

\begin{document}

\title{Uncertainty Quantification in LV State Estimation Under High Shares of Flexible Resources}

\author{
\IEEEauthorblockN{Nils~Müller, Samuel~Chevalier, Carsten~Heinrich, Kai~Heussen, Charalampos~Ziras}
\IEEEauthorblockA{\\Department of Electrical Engineering\\
Technical University of Denmark\\
Lyngby, Denmark\\
\{nilmu; schev; cahei; kh; chazi\}@elektro.dtu.dk}
}

\maketitle

\begin{abstract} 
The ongoing electrification introduces new challenges to \glspl{DSO}. 
Controllable resources may simultaneously react to price signals, potentially leading to network violations.
\Glspl{DSO} require reliable and accurate \gls{LVSE} to improve awareness and mitigate such events. 
However, the influence of flexibility activations on \gls{LVSE} has not been addressed yet.
It remains unclear if flexibility-induced uncertainty can be reliably quantified to enable robust \gls{DSO} decision-making.
In this work, uncertainty quantification in \gls{LVSE} is systematically investigated for multiple scenarios of input availability and flexibility utilization, using real data. 
For that purpose, a \gls{BNN} is compared to quantile regression. 
Results show that frequent flexibility activations can significantly deteriorate \gls{LVSE} performance, unless secondary substation measurements are available.
Moreover, it is demonstrated that the \gls{BNN} captures flexibility-induced voltage drops by dynamically extending the prediction interval during activation periods, and that it improves interpretability regarding the cause of uncertainty.
\end{abstract}


\IEEEpeerreviewmaketitle
\glsresetall
\section{Introduction}
\label{sec:Introduction}
In light of the European effort to reach carbon neutrality by $2050$, \glspl{DN} must undergo radical changes. 
Traditionally designed for the supply of consumers based on centralized generation, \glspl{DN} turn into carriers of volatile and often bidirectional power flows \cite{6395792}.
Key drivers are the increasing deployment of variable distributed generation, as well as the proliferation of \glspl{EV}, heat pumps and residential storage. 
To balance consumption and generation in renewable-based power systems, it is widely acknowledged that more local consumption flexibility is required \cite{hillberg2019flexibility}. 
However, controllable \glspl{DER} may react to price signals with a sudden change of power consumption, resulting in higher coincidence factors in \glspl{DN} dominated by such resources \cite{habib2018comprehensive}.
As a consequence, transformer or line protections could systematically be triggered, or unacceptably high voltage deviations occur.
Since large parts of the \gls{DN} are unobserved, \glspl{DSO} require techniques for reliable and accurate \gls{LVSE} in order to (i) improve awareness about the potential negative side effects of flexibility utilization and distributed generation, and (ii) be able to mitigate any potential operational problems.

For conventional \gls{SE}, network topology and line parameters must be known.
Moreover, since \gls{LVSE} constitutes a mathematically underdetermined problem, pseudo measurements are typically required to account for low meter coverage \cite{primadianto2016review}.
A widely considered approach to overcome these shortcomings is machine learning \cite{primadianto2016review,dehghanpour2018survey,hayes2014state}, whose feasibility and high estimation accuracy have been extensively demonstrated \cite{pertl2018validation,zamzam2020physics,menke2019distribution}.

A relatively small number of works have investigated the topic of probabilistic \gls{LVSE} so far.
Importantly, the influence of frequent flexibility activations on reliability and accuracy has not been sufficiently addressed.
Moreover, it remains unclear whether estimation uncertainty introduced by flexibility can be reliably quantified to support \gls{DSO} decision making, especially under varying levels of real-time data availability.

\subsection{Related work}
\label{subsec:Related_work}
A number of works propose the use of \glspl{NN} for real-time \gls{LVSE}.
For example, \cite{pertl2018validation} proposes and validates the use of a \gls{NN}, utilizing only real-time secondary substation information.
The application is seen as a real-time bus voltage estimator, with \gls{SM} data being available with a latency of one day. 
The authors of \cite{Ferdowsi2015} also propose the use of a \gls{NN} for estimating voltage magnitudes based on the same input, but do not use real data.
Further, the authors stressed that model performance is sufficient, largely because of the relatively simple topology that lacks long side branches, and the lack of significant amounts of \gls{PV} in-feed.

However, given the high variability of \gls{LV} network states and the increasing penetration of \glspl{DER}, a probabilistic approach seems more suitable compared to a deterministic output. 
The authors of \cite{ZUFFEREY2020106562} use quantile \glspl{NN} to provide probabilistic forecasts of the states of an \gls{LV} network.
However, a high degree of observability is assumed (known voltage and bus injections), and focus is given to the model's forecasting abilities.
Further, it is shown that \gls{EV} charging creates large uncertainty, which the authors mitigate by assuming full knowledge of the \gls{EV} charging start time and duration by the \gls{DSO}.

Ref. \cite{Bessa2018} proposes a data-driven probabilistic \gls{LVSE} based on an analog search technique and kernel density estimation.
This approach relies on finding similar past patterns, and it assumes real-time information from the secondary substation and voltage values from a number of \gls{LV} buses, while validation is performed by assessing the impact of varying levels of \gls{PV} penetration.
The authors of \cite{Sun2020} apply a variation of \gls{WLS} for probabilistic \gls{SE}.
However, a series of real-time measurements is assumed to be available, and hourly data is used without considerations for flexibility activation.

In \cite{mestav2019}, a deep learning approach to Bayesian \gls{SE} is proposed. 
The authors propose to first learn the distribution of bus injections from \gls{SM} data. 
Based on samples drawn from the learned distributions, a traditional feed-forward \gls{NN} is trained for the minimum mean-squared-error estimation of the system state.
Due to the application of a deterministic regression model, the proposed approach is not inherently probabilistic. 
The work does not consider uncertainty quantification, and it compares results only on deterministic error metrics. 
Moreover, no flexible resources are considered. 
The authors in \cite{Huang2019b} propose a deep belief network for pseudo measurements modeling. 
Based on an extended \gls{WLS} estimator, the probability density function of system states is inferred. 
However, the work assumes partial real-time knowledge of voltage states, and only \gls{MV} states are estimated. 

\subsection{Contribution and paper structure}
\label{subsec:Contribution_and_paper_structure}
The literature review shows that most works on \gls{LVSE} do not consider probabilistic approaches. 
Moreover, effects of varying levels of input information and flexibility usage on the estimation and uncertainty quantification performance are rarely studied. 
In this work, a \gls{BNN} is applied for probabilistic estimation of multiple \gls{LV} bus voltages. 
The \gls{BNN} is selected as it constitutes an inherently probabilistic approach, combining benefits of Bayesian uncertainty quantification and the predictive power of \glspl{NN}.
The main contributions of this work are as follows:
\begin{itemize}
    \item Systematic evaluation of the influence of different flexibility usage and input information scenarios on accuracy and uncertainty of \gls{LVSE}
    \item First application of a \gls{BNN} on \gls{LVSE} and comparison to a \gls{QR} benchmark
    \item First work considering and discussing epistemic and aleatoric uncertainty for probabilistic \gls{LVSE}.
\end{itemize}

The remainder of the paper is structured as follows. 
In Section \ref{sec:Method}, the use of a \gls{BNN} for \gls{LVSE} is motivated and a theoretical description of the applied model and its implementation provided. 
Section \ref{sec:Experimental_setup} presents the experimental setup, including the dataset, flexibility scenarios and \glspl{FS}, as well as the applied performance metrics. 
In Section \ref{sec:Results}, results are presented and discussed, followed by a conclusion and a view on future work in Section \ref{sec:Conclusion}. 

\section{Method} \label{sec:Method}
This section first motivates the selection of a \gls{BNN}. 
Next, the theoretical background and implementation are described.

\subsection{Model selection} \label{subsec:Model_selection}
Accurate \gls{SE} is the basis for the decision-making process of \glspl{DSO}. 
However, even an accurate estimator may result in large inaccuracies under specific or new circumstances, such as a certain time of the day or rare social events. 
Under these conditions, a deterministic estimator fails silently, affecting the \gls{DSO} decision-making process and potentially impacting critical decisions. 
In contrast, a probabilistic estimator can capture prediction uncertainty. 
By expressing \textit{what}, \textit{when} and \textit{why} it does not know, such an estimator increases interpretability of predictions. 
Thus, incorporating uncertainty quantification in \gls{LVSE} allows for risk-aware \glspl{DSO} network operation. 

Uncertainty can be classified into aleatoric (data) and epistemic (model). 
Aleatoric uncertainty is introduced by randomness in the process. 
In case of \gls{LVSE}, this randomness is given by factors such as measurement errors and random consumer behavior. 
Adding to this type of uncertainty, a given substation measurement can correspond to a variety of load realizations and thus \gls{LV} states, making the mapping of measurements to unobserved states non-unique.
Epistemic uncertainty comprises of model structure and parameter uncertainty due to lack of knowledge. 
In the \gls{LVSE} problem, this is given for example by data non-stationarity. 
If a model is only trained on winter months, it will encounter high epistemic uncertainty when predicting summer months, since input features are out-of-distribution. 
A fundamental difference between epistemic and aleatoric uncertainty is the fact that only the former can be reduced through additional information. 
To account for total uncertainty in \gls{LVSE}, a model capable of capturing both components is required. 

While quantifying estimation uncertainty is seen as an important requirement, accurate predictions are indispensable. 
As presented in Section \ref{subsec:Related_work}, multiple works have demonstrated the predictive power of \glspl{NN} for \gls{LVSE}. 
However, most of the available models are not able to represent uncertainty.

\glspl{BNN} constitute a new direction in machine learning \cite{jospin2020hands}. By connecting Bayesian statistics and deep learning, \glspl{BNN} combine the benefits of Bayesian uncertainty quantification with the predictive power of \glspl{NN}. 
In contrast to traditional \glspl{NN}, model parameters of \glspl{BNN} are not fixed. 
Instead, every weight and bias is represented by a conditional probability distribution, representing the uncertainty of the respective parameter. 
Predictions are generated through posterior inference. 
By directly sampling from the probabilistic parameters, \glspl{BNN} are inherently probabilistic, instead of deterministic, models.   

\subsection{BNN description} \label{subsec:BNN}
Let $X_{\text{train}}=\{x_{1},\dots,x_{N_{\text{train}}}\}$ and $Y_{\text{train}}=\{y_{1},\dots,y_{N_{\text{train}}}\}$ be the training input and output data, respectively, with $N_{\text{train}}$ being the number of training samples. 
The \gls{BNN} can be formulated as
\begin{equation}
[\hat{y},\hat{\sigma}^{2}] = f^{W}_{\text{BNN}}(x), \label{eq:BNN}
\end{equation}
where $W = \{W_{1},...,W_{N_{\text{L}}}\}$ are model parameters and $N_{\text{L}}$ the number of network layers. 
To consider aleatoric uncertainty, the output of the model is an estimate of both the predictive mean $\hat{y}$ and variance $\hat{\sigma}^{2}$. 
To account for epistemic uncertainty, a prior distribution is placed over $W$. 
In this work, a Gaussian prior $\mathcal{N}(0,I)$ is applied since Gaussian priors for \glspl{BNN} are known to provide the benefit of regularization \cite{vladimirova2019understanding}. 
The posterior  distribution $p(W|X_{\text{train}},Y_{\text{train}})$ over the model parameters, given the training data $\{X_{\text{train}}, Y_{\text{train}}\}$ is calculated by Bayes rule. 
The predictive distribution for a new observation $x$ is obtained by marginalizing over the posterior distribution \cite{wilson2020bayesian} according to
\begin{equation}
p(y|x,X_{\text{train}},Y_{\text{train}}) = \int p(y|x,W)p(W|X_{\text{train}},Y_{\text{train}})\text{d}W.
\end{equation}

Due to the non-linearity and non-conjugacy of \glspl{NN}, the true posterior is typically intractable. 
By minimizing the \gls{KL} divergence between $p(W|X_{\text{train}},Y_{\text{train}})$ and a surrogate distribution, the posterior is approximated. 
In this work, variational inference was used as inference algorithm for minimizing the \gls{KL} divergence \cite{hoffman2013stochastic}. 
To allow for simultaneous output of $\hat{y}$ and $\hat{\sigma}^{2}$ \eqref{eq:BNN}, and thus include aleatoric uncertainty, the loss function \cite{kendall2017uncertainties} of the \gls{BNN} is formulated as
\begin{equation}
\mathcal{L}_{\text{BNN}}(\theta) = \dfrac{1}{N_{\text{train}}}\sum^{N_{\text{train}}}_{i=1}\frac{1}{2\hat{\sigma}^{2}_{i}}||y_{i}-\hat{y}_{i}||^2+\frac{1}{2}\log \hat{\sigma}^{2}_{i}.
\end{equation}

To take epistemic uncertainty into account, multiple predictions of $\hat{y}$ and $\hat{\sigma}^{2}$ for input $x$ are required. Note that every prediction is based on a new set of sampled model parameters $\hat{W}_{t}$. The predictive mean is estimated with 
\begin{equation}
\widetilde{\mathbb{E}}(y)\approx\frac{1}{N_{\text{sample}}}\sum^{N_{\text{sample}}}_{t=1}f^{\hat{W}_{t}}_{\text{BNN}}(x),
\end{equation}
where $N_{\text{sample}}$ denotes the number of stochastic forward passes through the \gls{BNN}. The total predictive uncertainty, composed of an epistemic and aleatoric term, is approximated with
\begin{equation}
\begin{split}
\widetilde{\text{Var}}(y) \approx &\underbrace{\left[\frac{1}{N_{\text{sample}}}\sum^{N_{\text{sample}}}_{t=1}\hat{y}_{t}^{2}-\left(\frac{1}{N_{\text{sample}}}\sum^{N_{\text{sample}}}_{t=1}\hat{y}_{t}\right)^{2}\right]}_\text{epistemic} \\ &+\underbrace{\frac{1}{N_{\text{sample}}}\sum^{N_{\text{sample}}}_{t=1}\hat{\sigma}^{2}_{t}}_\text{aleatoric}.
\end{split}
\end{equation}

In this work, all investigated datasets are split into a training, validation and test set. 
A detailed explanation of the datasets, including input features and output variables, follows in Section \ref{sec:Experimental_setup}. 
Unless explicitly defined differently, the partition is $80/10/10$. 
The selection of model structure and hyperparameters is realized based on the validation loss. 
For all datasets, the smallest validation loss is achieved by using Adam optimizer, tanh activation function, two hidden layers and a batch size of $64$. 
The number of required epochs and units in the hidden layers varies in the range $2000$-$10000$ and $5$-$12$, respectively. 
The selected models are retrained on the respective training and validation data. 
Depending on the dataset and input features, the number of trainable parameters varies between $3780$ and $26766$.

\section{Experimental setup} \label{sec:Experimental_setup}
This section first describes the considered real network and load dataset, and explains the creation of the various flexibility usage and input information scenarios. 
Thereafter, metrics used for evaluating the \gls{LVSE} performance are introduced.

\subsection{Network and used dataset} \label{ssec:dataset}
The network used in this study is a real suburban \gls{MV}-\gls{LV} network located in Bornholm, Denmark, and is depicted in Fig. \ref{fig:net}. 
It consists of six \SI{10/0.4}{\kilo\volt} secondary substations, all connected to a single primary \SI{60/10}{\kilo\volt} substation, whose high voltage side is taken as the reference voltage and is set to $1$\,pu.
The network serves a total of $564$ residential customers.
A detailed model of the \gls{LV} network of secondary substation SubS is shown in Fig. \ref{fig:net}, while the remainder are modelled as PQ buses that represent the aggregated active and reactive power of all customers connected to the respective substation.

In this work, real $5$ minute average active and reactive power measurements from a large number of residential customers on Bornholm are used.
Data was collected during the EcoGrid~2.0 project \cite{ECO_HEINRICH} and also includes flexibility activation of heating loads, as a result of experimental demonstration of a local flexibility market \cite{ECO_ZIRAS, ECO_HEINRICH}.
Project participants are randomly assigned to the leaf nodes below SubS, and to the aggregated PQ profiles of the other secondary substations.
This approach was followed because during the project only a small portion of project participants were connected to the depicted network.
Time span of the study is the full year $2018$.

\subsection{Creating datasets for network states} \label{ssec:flow}
\gls{SM} data is assumed to become available to the \gls{DSO} with a daily delay.
Depending on infrastructure and \gls{DSO} practice, voltage values may also be extracted by \glspl{SM}.
If this is the case, a historical dataset can be created that contains all relevant voltage values.
If this information is unavailable, an accurate network model is needed to construct this dataset by using \gls{SM} consumption data and running power flows.

\begin{figure} [t]
  \centering
  \includegraphics[clip, trim=0.71cm 0.6cm 0.8cm 0.57cm,width=1\columnwidth]{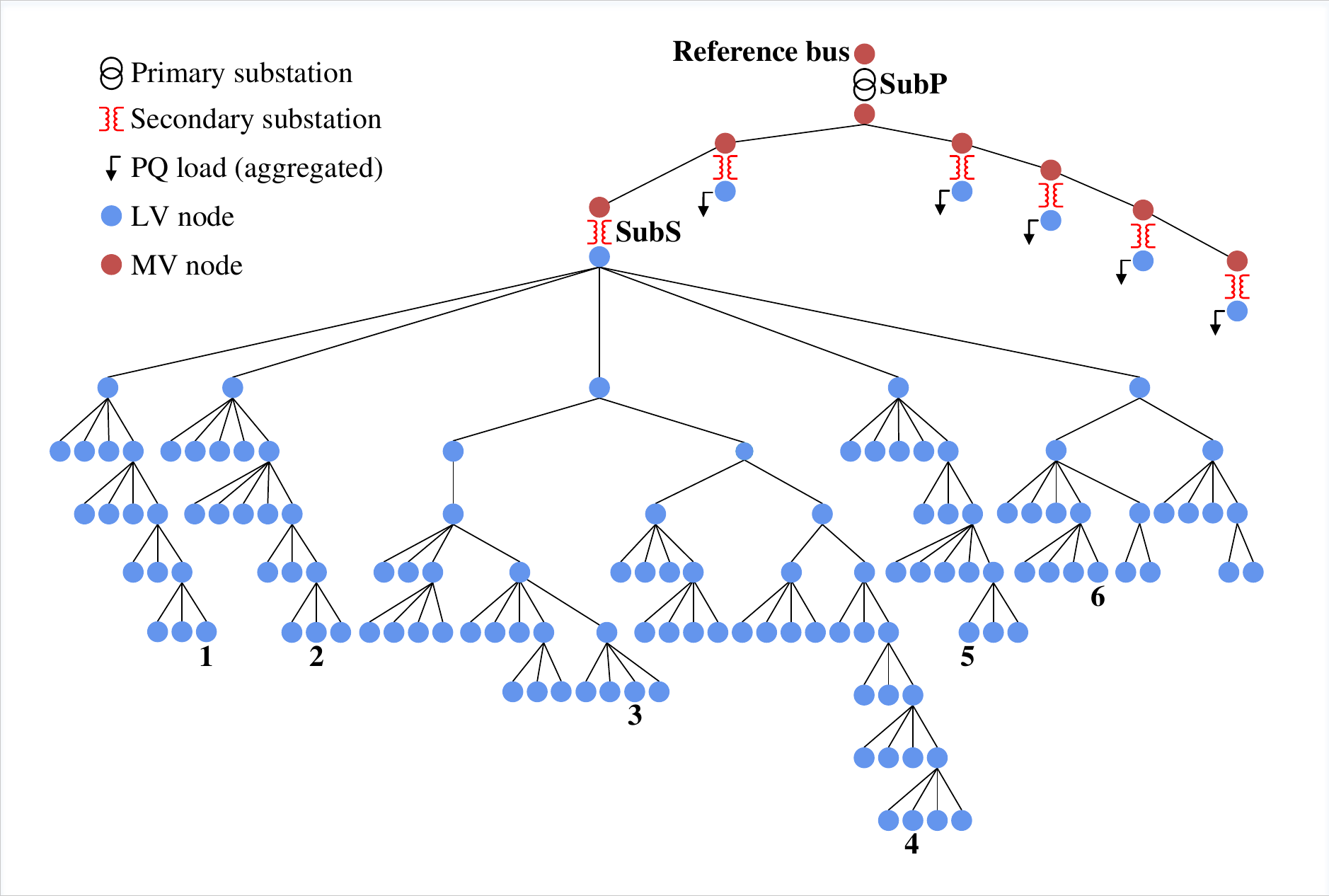}
\caption{Real \gls{MV}-\gls{LV} network used in the case study.} \label{fig:net}
\end{figure}

However, in most cases \glspl{DSO} have no real-time observability below the secondary substation level.
The objective is to provide a probabilistic estimation of such \gls{LV} states, and more specifically in this study, for voltages at nodes $1$ to $6$, marked on Fig. \ref{fig:net}, given different flexibility utilization scenarios and varying levels of information availability to the \gls{DSO}.
The network model and its accuracy have been validated with real network measurements. 
Since no voltage measurements are available from the \glspl{SM}, AC power flow was used to obtain the network states, which serve as ground truth.

\subsection{Flexibility scenarios} \label{ssec:scenarios}
\subsubsection{S1: original data}
This scenario considers the original residential profiles dataset, and customers are assigned randomly to the leaf nodes of the \gls{LV} grid and the other five secondary substations.
S1 presents a case with mild flexibility utilization.
\subsubsection{S2: original data and EVs below adjacent substations}
In this scenario the customer placement of S1 is kept, but \glspl{EV} are added on the five adjacent secondary substations of SubS.
\Gls{EV} profiles are taken from \cite{EVdata} and it is assumed that users perform smart charging with the objective of minimizing their costs based on retail prices, which follow the day-ahead spot prices of Danish zone DK2.
S2 presents a case with significant use of flexibility, which potentially adds randomness and reduces the correlation between the considered \gls{LV} nodes and the primary substation.
\subsubsection{S3: original data and EVs below all substations}
In the third scenario, an \gls{EV} performing smart charging also is added to each customer under SubS. 
S3 presents a case with high flexibility-induced load variability in the examined network below SubS, potentially complicating voltage estimation. 

\subsection{DSO information availability} \label{ssec:information}
Observability in \gls{LV} networks is still rather limited. This work assesses how different levels of \gls{DSO} information availability affect the performance of probabilistic \gls{SE}, for each of the aforementioned three flexibility scenarios.
Below three \glspl{FS} with increasing data availability are described.

\subsubsection{FS1: low availability}
The \gls{DSO} has access to past customer \gls{SM} data (with a delay of $24$ hours), weather data and calendric features (like weekday vs weekend indicators and time of the day).
Additionally, the retailer prices are used, which is beneficial for scenarios S2 and S3 (smart charging).
\gls{FS}1 assumes zero \gls{DSO} real-time observability and the used information can be readily available to every \gls{DSO} without the installation of additional devices.
\subsubsection{FS2: medium availability}
In \gls{FS}2 it is assumed that real-time measurements from SubP are available, which is not an unusual operational practice. 
Therefore, primary substation PQ values are considered as additional features on top of \gls{FS}1 to construct \gls{FS}2.
Note that adding voltage values from SubP was found to add a negligible benefit to the performance of the models and was thus omitted. 

\subsubsection{FS3: high availability}
The high availability case \gls{FS}3 assumes that the \gls{DSO} also monitors in real time the secondary substation and thus PQ and voltage values from SubS are added as features on top of the ones from \gls{FS}2.

\subsection{Evaluation} \label{ssec:metrics}
In this work, the \gls{BNN} is benchmarked with \gls{QR}, which is a straightforward method to conduct probabilistic state estimation that showed good performance in many applications \cite{koenker2017handbook}.
The performance of the models is evaluated for each flexibility scenario (S1, S2 and S3) and each \gls{FS} (\gls{FS}1, \gls{FS}2 and \gls{FS}3) on the test dataset. For evaluation of the point estimation performance
\gls{RMSE} is used. The popular Pinball and Winkler scores are considered for assessing the reliability, sharpness, and resolution of the probabilistic estimation.
These metrics are calculated for each considered network state $j \in \mathcal{J}$ individually. In Section \ref{sec:Results} their average, minimum and maximum value are reported.
$y_{j,t}$ is the actual value of network state $j$ at the evaluated time step $t$ (out of $n$ steps).
$\hat{y}^\text{m}_{j,t}$ is the expectation of the predicted value. Note that this expectation corresponds to the median for \gls{QR} and the mean for the \gls{BNN}.
Finally, $\hat{y}^q_{j,t}$ is the predicted $q-$th quantile.
\Gls{RMSE} is defined as
\begin{align} \label{eq:RMSE}
RMSE_j = \sqrt{\frac{\sum_{t}^{n}\left (y_{j,t}-\hat{y}^\text{m}_{j,t}  \right )^2}{n}}, ~~~~ \forall j \in \mathcal{J}.
\end{align}

The Pinball loss function is given by
\begin{equation} \label{eq:Pinball}
Pinball_j = \begin{cases}
(y_{j,t}-\hat{y}^q_{j,t}) q, & y_{j,t}\geq\hat{y}^q_{j,t}\\ 
(\hat{y}^q_{j,t}-y_{j,t}) (1-q), & y_{j,t}<\hat{y}^q_{j,t}.
\end{cases}
\end{equation}

The average Pinball score for all $n$ steps is calculated for $q = 0.01,\dots,0.99$, with a lower value indicating better performance.
Finally, Winkler score for a \gls{PI} $1-\alpha$ is given by
\begin{equation} \label{eq:Winkler}
Winkler_j = \begin{cases}
\delta, & \hat{y}^-_{j,t} \leq y_{j,t} \leq \hat{y}^+_{j,t}\\
2(\hat{y}^-_{j,t} - y_{j,t})/\alpha + \delta, & y_{j,t} < \hat{y}^-_{j,t}\\ 
2(y_{j,t} - \hat{y}^+_{j,t})/\alpha + \delta, & y_{j,t} > \hat{y}^+_{j,t},
\end{cases}
\end{equation}
where $\hat{y}^-_{j,t}$ and $\hat{y}^+_{j,t}$ represent the lower and upper \gls{PI} bounds, respectively. 
$\delta = \hat{y}^+_{j,t} - \hat{y}^-_{j,t}$ and $\alpha = 0.1$ because a \SI{90}{\percent} \gls{PI} is considered.
A lower Winkler score implies a better \gls{PI}.
\section{Results} \label{sec:Results}
In this section the voltage estimation performance of the \gls{BNN} under varying input feature and flexibility scenarios is evaluated and compared to a \gls{QR} benchmark. In Subsection \ref{subsec:Qualitative_assessment} a qualitative evaluation is presented, followed by a quantitative assessment in Subsection \ref{subsec:Quantitative_assessment}. In Subsection \ref{subsec:uncertainty_evaluation} the behavior of aleatoric and epistemic uncertainty is evaluated over a period of $10$ months. 

\subsection{Qualitative evaluation of the voltage estimation} \label{subsec:Qualitative_assessment}
\begin{figure}[t]
\centering
    \input{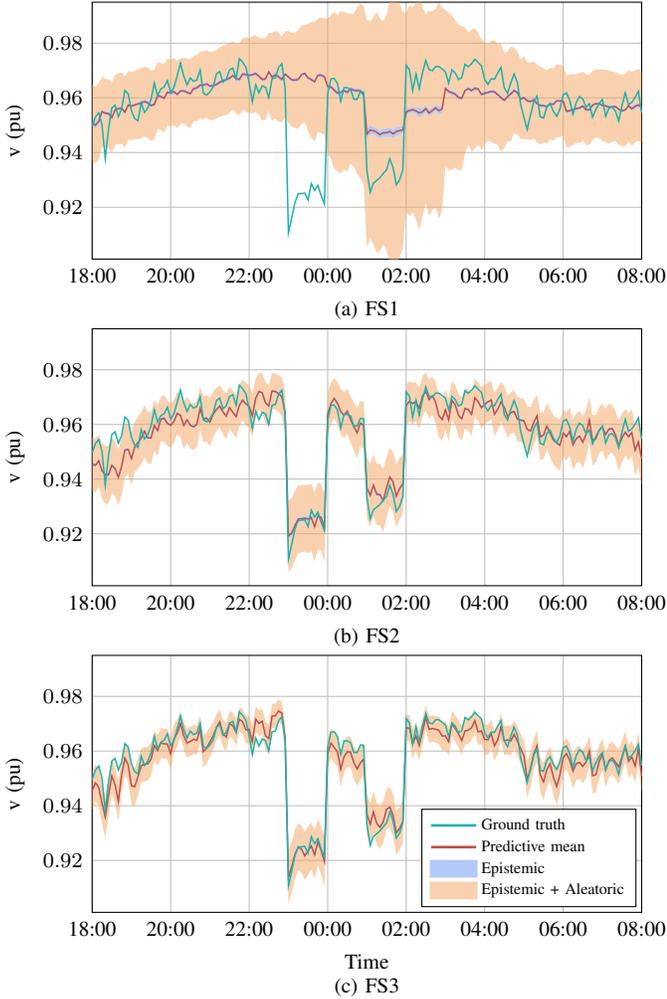}
    \caption{Comparison of the predictive mean and \SI{90}{\percent} \gls{PI} for \gls{FS}1-\gls{FS}3 based on a representative excerpt from S3 and bus voltage 4. Selection of bus voltage 4 results from proximity to the average \gls{RMSE} of all predicted bus voltages.} \label{fig:FScomparison}
\end{figure}
Fig. \ref{fig:FScomparison} shows the output of the \gls{BNN} for a $14$ hours period of bus voltage $4$ under three levels of available information (\gls{FS}1-\gls{FS}3).
Case S3 is considered, where substantial flexibility activations occur in the examined network below SubS.
This is evident during periods $23$:$00$-$00$:$00$ and $01$:$00$-$02$:$00$, where large load increases occur due to \gls{EV} charging, leading to low voltage values.
For all \glspl{FS}, aleatoric uncertainty is dominating, which can be explained by the comprehensive training dataset.
However, an increase of the epistemic uncertainty can be noticed during flexibility activations. 
This can be explained by a lack of knowledge, as they constitute only a small subset of the training data. 
For all \glspl{FS} the \gls{BNN} understands that uncertainty during flexibility activations is higher, resulting in an extended \gls{PI}. 
However, for \gls{FS}1 the model entirely misses the first voltage drop. 
Under \gls{FS}1, no real-time network states are considered. 
It can be concluded that \gls{FS}1 is not sufficient to estimate flexibility-driven voltage drops. 
By incorporating primary (\gls{FS}2) or secondary (\gls{FS}3) substation measurements the model provides \glspl{PI} which successfully capture voltage drops, as the increased consumption which causes them is reflected on the substation's power and voltage. An additional advantage of including secondary substation measurements can be seen from the tighter \gls{PI}. 
\begin{figure}[t]
\centering
    \input{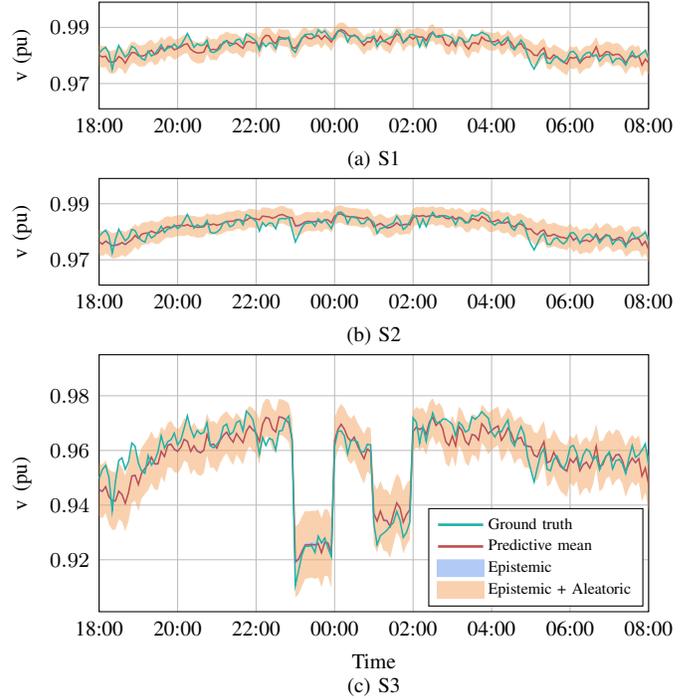}
    \caption{Comparison of the predictive mean and \SI{90}{\percent} \gls{PI} for a representative excerpt of S1-S3 based on \gls{FS}2 and bus voltage 4. Selection of bus voltage 4 results from proximity to the average \gls{RMSE} of all predicted bus voltages.} \label{fig:Scomparison}
\end{figure}

Next, only \gls{FS}2 is considered, which is the typical scenario for most \glspl{DSO}. 
In Fig. \ref{fig:Scomparison}, results are depicted for the same period under the three flexibility usage scenarios (S1-S3). 
In all scenarios epistemic uncertainty is almost entirely reduced by the \gls{BNN} learning process. 
This leads to the conclusion that $11$ months of training data is sufficient to exploit available information. 
Remaining uncertainty is mainly introduced by randomness. 
In S1 load variability and flexibility use are limited. 
Thus, the \gls{BNN} is able to provide good voltage estimation with a very tight \gls{PI}. 
Similar results are obtained under S2. 
Visual inspection let assume that frequent flexibility activations below adjacent substations have small impact on the estimation. 
A quantitative evaluation follows in Subsection \ref{subsec:Quantitative_assessment}. 
For S3 substantial flexibility is active in the examined \gls{LV} network. 
The wider \gls{PI} indicates that estimations are less accurate. 
Moreover, an increase of the \gls{PI} is noticed during flexibility activations. 
The high \gls{PI} resolution allows the \gls{BNN} to successfully capture the voltage drops.

\subsection{Quantitative evaluation and benchmark comparison} \label{subsec:Quantitative_assessment}
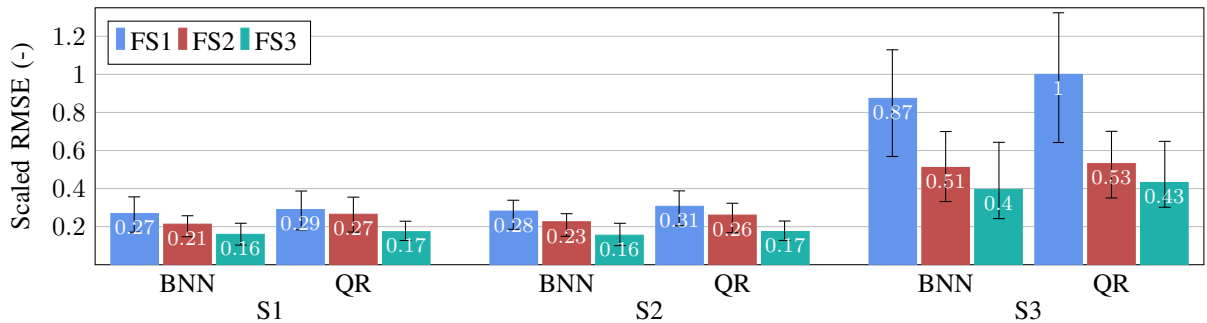
\begin{figure*}[t]
\centering
\begin{tikzpicture}
    \begin{axis}[
        width  = 0.9\textwidth, 
        height = 5cm, 
        major x tick style = transparent,
        major y tick style = transparent,
        ybar=2pt,
        bar width=18pt, 
        nodes near coords,
        every node near coord/.append style={yshift=-0.4cm},
        every node near coord/.append style={color=white},
        every node near coord/.append style={font=\footnotesize},
        ymajorgrids = true,
        xtick = {1.2,2,2.8},
        ytick = {0.2, 0.4, 0.6, 0.8, 1, 1.2},
        xticklabels={S1, S2, S3},
        xticklabel style={
        yshift=-6pt, 
        },
        extra x ticks={1.025,1.375,1.825,2.175, 2.625, 2.975},
        extra x tick labels={BNN, QR, BNN, QR, BNN, QR},
        extra x tick style={
        yshift=6pt,
        tickwidth=0},
        ymax=1.35,
        ymin=0,
        ylabel = {Scaled RMSE (-)},
        xlabel style={yshift=7pt},
        legend pos=north west,
        legend style={
               at={(0.103,0.95)},
                anchor=north,
                column sep=0ex,
               legend columns=5
        }
    ]
        \addplot[style={cornflowerblue,fill=cornflowerblue,mark=none}] 
            coordinates {(1.025, 0.0028499/0.0106083) (1.375, 0.0030735/0.0106083) (1.825, 0.0029905/0.0106083) (2.175, 0.0032542/0.0106083) (2.625, 0.0092703/0.0106083) (2.975, 0.0106083/0.0106083)};

        \addplot[style={rred,fill=rred,mark=none}]
            coordinates {(1.025, 0.0022598/0.0106083) (1.375, 0.0028122/0.0106083) (1.825, 0.0023936/0.0106083) (2.175, 0.0027651/0.0106083) (2.625, 0.0054205/0.0106083) (2.975, 0.0056364/0.0106083)};

        \addplot[style={lightseagreen,fill=lightseagreen,mark=none}]
            coordinates {(1.025, 0.0016919/0.0106083) (1.375, 0.0018428/0.0106083) (1.825, 0.0016444/0.0106083) (2.175, 0.0018512/0.0106083) (2.625, 0.0042028/0.0106083) (2.975, 0.0045826/0.0106083)};
            
        \draw [solid] (0.9125,0.0018168/0.0106083) -- (0.9125,0.003782/0.0106083);
        \draw [solid] (0.9,0.003782/0.0106083) -- (0.925,0.003782/0.0106083);
        \draw [solid] (0.9,0.0018168/0.0106083) -- (0.925,0.0018168/0.0106083);
        
        \draw [solid] (1.025,0.0015618/0.0106083) -- (1.025,0.0027292/0.0106083);
        \draw [solid] (1.0125,0.0027292/0.0106083) -- (1.0375,0.0027292/0.0106083);
        \draw [solid] (1.0125,0.0015618/0.0106083) -- (1.0375,0.0015618/0.0106083);
        
        \draw [solid] (1.1375,0.0010912/0.0106083) -- (1.1375,0.002311/0.0106083);
        \draw [solid] (1.125,0.002311/0.0106083) -- (1.15,0.002311/0.0106083);
        \draw [solid] (1.125,0.0010912/0.0106083) -- (1.15,0.0010912/0.0106083);

        \draw [solid] (1.265,0.0019429/0.0106083) -- (1.265,0.004108/0.0106083);
        \draw [solid] (1.2525,0.004108/0.0106083) -- (1.2775,0.004108/0.0106083);
        \draw [solid] (1.2525,0.0019429/0.0106083) -- (1.2775,0.0019429/0.0106083);
        
        \draw [solid] (1.375,0.001819/0.0106083) -- (1.375,0.003766/0.0106083);
        \draw [solid] (1.3625,0.003766/0.0106083) -- (1.3875,0.003766/0.0106083);
        \draw [solid] (1.3625,0.001819/0.0106083) -- (1.3875,0.001819/0.0106083);
        
        \draw [solid] (1.485,0.0013509/0.0106083) -- (1.485,0.0024251/0.0106083);
        \draw [solid] (1.4725,0.0024251/0.0106083) -- (1.4975,0.0024251/0.0106083);
        \draw [solid] (1.4725,0.0013509/0.0106083) -- (1.4975,0.0013509/0.0106083);
        
        \draw [solid] (1.7125,0.0019721/0.0106083) -- (1.7125,0.0035933/0.0106083);
        \draw [solid] (1.7,0.0035933/0.0106083) -- (1.725,0.0035933/0.0106083);
        \draw [solid] (1.7,0.0019721/0.0106083) -- (1.725,0.0019721/0.0106083);
        
        \draw [solid] (1.825,0.0015808/0.0106083) -- (1.825,0.0028438/0.0106083);
        \draw [solid] (1.8125,0.0028438/0.0106083) -- (1.8375,0.0028438/0.0106083);
        \draw [solid] (1.8125,0.0015808/0.0106083) -- (1.8375,0.0015808/0.0106083);
        
        \draw [solid] (1.9375,0.0010627/0.0106083) -- (1.9375,0.0023101/0.0106083);
        \draw [solid] (1.925,0.0023101/0.0106083) -- (1.95,0.0023101/0.0106083);
        \draw [solid] (1.925,0.0010627/0.0106083) -- (1.95,0.0010627/0.0106083);
        
        \draw [solid] (2.0625,0.0021718/0.0106083) -- (2.0625,0.004116/0.0106083);
        \draw [solid] (2.05,0.004116/0.0106083) -- (2.075,0.004116/0.0106083);
        \draw [solid] (2.05,0.0021718/0.0106083) -- (2.075,0.0021718/0.0106083);    
        
        \draw [solid] (2.175,0.001791/0.0106083) -- (2.175,0.003426/0.0106083);
        \draw [solid] (2.1625,0.003426/0.0106083) -- (2.1875,0.003426/0.0106083);
        \draw [solid] (2.1625,0.001791/0.0106083) -- (2.1875,0.001791/0.0106083);
        
        \draw [solid] (2.285,0.0013537/0.0106083) -- (2.285,0.0024353/0.0106083);
        \draw [solid] (2.2725,0.0024353/0.0106083) -- (2.2975,0.0024353/0.0106083);
        \draw [solid] (2.2725,0.0013537/0.0106083) -- (2.2975,0.0013537/0.0106083);
        
        \draw [solid] (2.5125,0.0060321/0.0106083) -- (2.5125,0.0119796/0.0106083);
        \draw [solid] (2.5,0.0119796/0.0106083) -- (2.525,0.0119796/0.0106083);
        \draw [solid] (2.5,0.0060321/0.0106083) -- (2.525,0.0060321/0.0106083);

        \draw [solid] (2.625,0.0035194/0.0106083) -- (2.625,0.0074224/0.0106083);
        \draw [solid] (2.6125,0.0074224/0.0106083) -- (2.6375,0.0074224/0.0106083);
        \draw [solid] (2.6125,0.0035194/0.0106083) -- (2.6375,0.0035194/0.0106083);
        
        \draw [solid] (2.7375,0.0025697/0.0106083) -- (2.7375,0.006822/0.0106083);
        \draw [solid] (2.725,0.006822/0.0106083) -- (2.75,0.006822/0.0106083);
        \draw [solid] (2.725,0.0025697/0.0106083) -- (2.75,0.0025697/0.0106083);
    
        \draw [solid] (2.86375,0.006811/0.0106083) -- (2.86375,0.0140428/0.0106083);
        \draw [solid] (2.85125,0.0140428/0.0106083) -- (2.875,0.0140428/0.0106083);
        \draw [solid] (2.85125,0.006811/0.0106083) -- (2.875,0.006811/0.0106083);    

        \draw [solid] (2.975,0.0037196/0.0106083) -- (2.975,0.0074357/0.0106083);
        \draw [solid] (2.9625,0.0074357/0.0106083) -- (2.9875,0.0074357/0.0106083);
        \draw [solid] (2.9625,0.0037196/0.0106083) -- (2.9875,0.0037196/0.0106083);
        
        \draw [solid] (3.0875,0.0032027/0.0106083) -- (3.0875,0.006875/0.0106083);
        \draw [solid] (3.1,0.006875/0.0106083) -- (3.075,0.006875/0.0106083);
        \draw [solid] (3.1,0.0032027/0.0106083) -- (3.075,0.0032027/0.0106083);
    
        \legend{\gls{FS}1, \gls{FS}2, \gls{FS}3}

    \end{axis}
\end{tikzpicture}
\caption{Scaled \gls{RMSE} of the \gls{BNN} and \gls{QR} for S1-S3 and \gls{FS}1-\gls{FS}3 averaged over all estimated bus voltages. Min/max values are indicated by the black bars.} \label{fig:RMSE}
\end{figure*}
\begin{figure*}[b]
\centering
\hspace*{-0.09cm}\vspace*{-0.2cm}
\begin{tikzpicture}
    \begin{axis}[xticklabels={,,},
        width  = 0.9\textwidth, 
        height = 5cm, 
        major x tick style = transparent,
        major y tick style = transparent,
        ybar=2pt,
        bar width=18pt, 
        nodes near coords,
        every node near coord/.append style={yshift=-0.4cm},
        every node near coord/.append style={color=white},
        every node near coord/.append style={font=\footnotesize},
        ymajorgrids = true,
        ytick = {0.2, 0.4, 0.6, 0.8, 1, 1.2},
        xticklabel style={
        yshift=-8pt, 
        },
        ymax=1.05,
        ymin=0,
        ylabel = {Scaled Pinball (-)},
        xlabel = {(a) Pinball},
        xlabel style={yshift=18pt},
        legend style={
               at={(0.9,0.198)},
                anchor=north,
                column sep=0ex,
               legend columns=5
        }
    ]
        \addplot[style={cornflowerblue,fill=cornflowerblue,mark=none}] 
            coordinates {(1.025, 0.000871/0.0032765) (1.375, 0.0009501/0.0032765) (1.825, 0.0008996/0.0032765) (2.175, 0.0009888/0.0032765) (2.625, 0.0028335/0.0032765) (2.975, 0.0032765/0.0032765)};

        \addplot[style={rred,fill=rred,mark=none}]
            coordinates {(1.025, 0.0006876/0.0032765) (1.375, 0.0008289/0.0032765) (1.825, 0.0007188/0.0032765) (2.175, 0.0008432/0.0032765) (2.625, 0.0015951/0.0032765) (2.975, 0.001716/0.0032765)};

        \addplot[style={lightseagreen,fill=lightseagreen,mark=none}]
            coordinates {(1.025, 0.0003739/0.0032765) (1.375, 0.0005092/0.0032765) (1.825, 0.000355/0.0032765) (2.175, 0.0005117/0.0032765) (2.625, 0.0008904/0.0032765) (2.975, 0.0012293/0.0032765)};
            
        \draw [solid] (0.9125,0.0008159/0.0032765) -- (0.9125,0.0008907/0.0032765);
        \draw [solid] (0.9,0.0008907/0.0032765) -- (0.925,0.0008907/0.0032765);
        \draw [solid] (0.9,0.0008159/0.0032765) -- (0.925,0.0008159/0.0032765);
        
        \draw [solid] (1.025,0.0006342/0.0032765) -- (1.025,0.0007072/0.0032765);
        \draw [solid] (1.0125,0.0007072/0.0032765) -- (1.0375,0.0007072/0.0032765);
        \draw [solid] (1.0125,0.0006342/0.0032765) -- (1.0375,0.0006342/0.0032765);
        
        \draw [solid] (1.1375,0.0002536/0.0032765) -- (1.1375,0.0004386/0.0032765);
        \draw [solid] (1.125,0.0004386/0.0032765) -- (1.15,0.0004386/0.0032765);
        \draw [solid] (1.125,0.0002536/0.0032765) -- (1.15,0.0002536/0.0032765);

        \draw [solid] (1.265,0.0008445/0.0032765) -- (1.265,0.0009874/0.0032765);
        \draw [solid] (1.2525,0.0009874/0.0032765) -- (1.2775,0.0009874/0.0032765);
        \draw [solid] (1.2525,0.0008445/0.0032765) -- (1.2775,0.0008445/0.0032765);
        
        \draw [solid] (1.375,0.0007273/0.0032765) -- (1.375,0.0008642/0.0032765);
        \draw [solid] (1.3625,0.0008642/0.0032765) -- (1.3875,0.0008642/0.0032765);
        \draw [solid] (1.3625,0.0007273/0.0032765) -- (1.3875,0.0007273/0.0032765);
        
        \draw [solid] (1.485,0.0004417/0.0032765) -- (1.485,0.0005385/0.0032765);
        \draw [solid] (1.4725,0.0005385/0.0032765) -- (1.4975,0.0005385/0.0032765);
        \draw [solid] (1.4725,0.0004417/0.0032765) -- (1.4975,0.0004417/0.0032765);
        
        \draw [solid] (1.7125,0.000832/0.0032765) -- (1.7125,0.0009229/0.0032765);
        \draw [solid] (1.7,0.0009229/0.0032765) -- (1.725,0.0009229/0.0032765);
        \draw [solid] (1.7,0.000832/0.0032765) -- (1.725,0.000832/0.0032765);
        
        \draw [solid] (1.825,0.0006587/0.0032765) -- (1.825,0.0007401/0.0032765);
        \draw [solid] (1.8125,0.0007401/0.0032765) -- (1.8375,0.0007401/0.0032765);
        \draw [solid] (1.8125,0.0006587/0.0032765) -- (1.8375,0.0006587/0.0032765);
        
        \draw [solid] (1.9375,0.0002361/0.0032765) -- (1.9375,0.0004201/0.0032765);
        \draw [solid] (1.925,0.0004201/0.0032765) -- (1.95,0.0004201/0.0032765);
        \draw [solid] (1.925,0.0002361/0.0032765) -- (1.95,0.0002361/0.0032765);
        
        \draw [solid] (2.0625,0.00089/0.0032765) -- (2.0625,0.0010229/0.0032765);
        \draw [solid] (2.05,0.0010229/0.0032765) -- (2.075,0.0010229/0.0032765);
        \draw [solid] (2.05,0.00089/0.0032765) -- (2.075,0.00089/0.0032765);    
        
        \draw [solid] (2.175,0.0007661/0.0032765) -- (2.175,0.0008703/0.0032765);
        \draw [solid] (2.1625,0.0008703/0.0032765) -- (2.1875,0.0008703/0.0032765);
        \draw [solid] (2.1625,0.0007661/0.0032765) -- (2.1875,0.0007661/0.0032765);
        
        \draw [solid] (2.285,0.0004427/0.0032765) -- (2.285,0.0005415/0.0032765);
        \draw [solid] (2.2725,0.0005415/0.0032765) -- (2.2975,0.0005415/0.0032765);
        \draw [solid] (2.2725,0.0004427/0.0032765) -- (2.2975,0.0004427/0.0032765);
        
        \draw [solid] (2.5125,0.0027645/0.0032765) -- (2.5125,0.0028852/0.0032765);
        \draw [solid] (2.5,0.0028852/0.0032765) -- (2.525,0.0028852/0.0032765);
        \draw [solid] (2.5,0.0027645/0.0032765) -- (2.525,0.0027645/0.0032765);

        \draw [solid] (2.625,0.001451/0.0032765) -- (2.625,0.0016577/0.0032765);
        \draw [solid] (2.6125,0.0016577/0.0032765) -- (2.6375,0.0016577/0.0032765);
        \draw [solid] (2.6125,0.001451/0.0032765) -- (2.6375,0.001451/0.0032765);
        
        \draw [solid] (2.7375,0.0005682/0.0032765) -- (2.7375,0.0010752/0.0032765);
        \draw [solid] (2.725,0.0010752/0.0032765) -- (2.75,0.0010752/0.0032765);
        \draw [solid] (2.725,0.0005682/0.0032765) -- (2.75,0.0005682/0.0032765);
    
        \draw [solid] (2.86375,0.0031953/0.0032765) -- (2.86375,0.003344/0.0032765);
        \draw [solid] (2.85125,0.003344/0.0032765) -- (2.875,0.003344/0.0032765);
        \draw [solid] (2.85125,0.0031953/0.0032765) -- (2.875,0.0031953/0.0032765);    

        \draw [solid] (2.975,0.0015435/0.0032765) -- (2.975,0.0017833/0.0032765);
        \draw [solid] (2.9625,0.0017833/0.0032765) -- (2.9875,0.0017833/0.0032765);
        \draw [solid] (2.9625,0.0015435/0.0032765) -- (2.9875,0.0015435/0.0032765);
        
        \draw [solid] (3.0875,0.0010354/0.0032765) -- (3.0875,0.0013216/0.0032765);
        \draw [solid] (3.1,0.0013216/0.0032765) -- (3.075,0.0013216/0.0032765);
        \draw [solid] (3.1,0.0010354/0.0032765) -- (3.075,0.0010354/0.0032765);
    

    \end{axis}
\end{tikzpicture}
\begin{tikzpicture}
    \begin{axis}[
        width  = 0.9\textwidth, 
        height = 5cm, 
        major x tick style = transparent,
        major y tick style = transparent,
        ybar=2pt,
        bar width=18pt, 
        nodes near coords,
        every node near coord/.append style={yshift=-0.4cm},
        every node near coord/.append style={color=white},
        every node near coord/.append style={font=\footnotesize},
        ymajorgrids = true,
        xtick = {1.2,2,2.8},
        ytick = {0.2, 0.4, 0.6, 0.8, 1, 1.2},
        xticklabels={S1, S2, S3},
        xticklabel style={
        yshift=-6pt, 
        },
        extra x ticks={1.025,1.375,1.825,2.175, 2.625, 2.975},
        extra x tick labels={BNN, QR, BNN, QR, BNN, QR},
        extra x tick style={
        yshift=6pt,
        tickwidth=0},
        ymax=1.35,
        ymin=0,
        ylabel = {Scaled Winkler (\SI{90}{\percent}) (-)},
        xlabel = {(b) Winkler (\SI{90}{\percent})},
        xlabel style={yshift=7pt},
        legend pos=north west,
        legend style={
               at={(0.103,0.95)},
                anchor=north,
                column sep=0ex,
               legend columns=5
        }
    ]
        \addplot[style={cornflowerblue,fill=cornflowerblue,mark=none}] 
            coordinates {(1.025, 0.0117066/0.0477508) (1.375, 0.0141634/0.0477508) (1.825, 0.012429/0.0477508) (2.175, 0.0142308/0.0477508) (2.625, 0.0386492/0.0477508) (2.975, 0.0477508/0.0477508)};

        \addplot[style={rred,fill=rred,mark=none}]
            coordinates {(1.025, 0.0091614/0.0477508) (1.375, 0.0112485/0.0477508) (1.825, 0.0099769/0.0477508) (2.175, 0.0118262/0.0477508) (2.625, 0.0229072/0.0477508) (2.975, 0.0279078/0.0477508)};

        \addplot[style={lightseagreen,fill=lightseagreen,mark=none}]
            coordinates {(1.025, 0.0069795/0.0477508) (1.375, 0.0074632/0.0477508) (1.825, 0.0068712/0.0477508) (2.175, 0.0074896/0.0477508) (2.625, 0.0178156/0.0477508) (2.975, 0.0186606/0.0477508)};
            
        \draw [solid] (0.9125,0.0077324/0.0477508) -- (0.9125,0.0137421/0.0477508);
        \draw [solid] (0.9,0.0137421/0.0477508) -- (0.925,0.0137421/0.0477508);
        \draw [solid] (0.9,0.0077324/0.0477508) -- (0.925,0.0077324/0.0477508);
        
        \draw [solid] (1.025,0.0064726/0.0477508) -- (1.025,0.010829/0.0477508);
        \draw [solid] (1.0125,0.010829/0.0477508) -- (1.0375,0.010829/0.0477508);
        \draw [solid] (1.0125,0.0064726/0.0477508) -- (1.0375,0.0064726/0.0477508);
        
        \draw [solid] (1.1375,0.004418/0.0477508) -- (1.1375,0.0097062/0.0477508);
        \draw [solid] (1.125,0.0097062/0.0477508) -- (1.15,0.0097062/0.0477508);
        \draw [solid] (1.125,0.004418/0.0477508) -- (1.15,0.004418/0.0477508);

        \draw [solid] (1.265,0.0089676/0.0477508) -- (1.265,0.0190949/0.0477508);
        \draw [solid] (1.2525,0.0190949/0.0477508) -- (1.2775,0.0190949/0.0477508);
        \draw [solid] (1.2525,0.0089676/0.0477508) -- (1.2775,0.0089676/0.0477508);
        
        \draw [solid] (1.375,0.0075674/0.0477508) -- (1.375,0.0143183/0.0477508);
        \draw [solid] (1.3625,0.0143183/0.0477508) -- (1.3875,0.0143183/0.0477508);
        \draw [solid] (1.3625,0.0075674/0.0477508) -- (1.3875,0.0075674/0.0477508);
        
        \draw [solid] (1.485,0.0054925/0.0477508) -- (1.485,0.0098698/0.0477508);
        \draw [solid] (1.4725,0.0098698/0.0477508) -- (1.4975,0.0098698/0.0477508);
        \draw [solid] (1.4725,0.0054925/0.0477508) -- (1.4975,0.0054925/0.0477508);
        
        \draw [solid] (1.7125,0.0085135/0.0477508) -- (1.7125,0.0144766/0.0477508);
        \draw [solid] (1.7,0.0144766/0.0477508) -- (1.725,0.0144766/0.0477508);
        \draw [solid] (1.7,0.0085135/0.0477508) -- (1.725,0.0085135/0.0477508);
        
        \draw [solid] (1.825,0.0066601/0.0477508) -- (1.825,0.0115249/0.0477508);
        \draw [solid] (1.8125,0.0115249/0.0477508) -- (1.8375,0.0115249/0.0477508);
        \draw [solid] (1.8125,0.0066601/0.0477508) -- (1.8375,0.0066601/0.0477508);
        
        \draw [solid] (1.9375,0.0044336/0.0477508) -- (1.9375,0.0098694/0.0477508);
        \draw [solid] (1.925,0.0098694/0.0477508) -- (1.95,0.0098694/0.0477508);
        \draw [solid] (1.925,0.0044336/0.0477508) -- (1.95,0.0044336/0.0477508);
        
        \draw [solid] (2.0625,0.0095257/0.0477508) -- (2.0625,0.0179626/0.0477508);
        \draw [solid] (2.05,0.0179626/0.0477508) -- (2.075,0.0179626/0.0477508);
        \draw [solid] (2.05,0.0095257/0.0477508) -- (2.075,0.0095257/0.0477508);    
        
        \draw [solid] (2.175,0.0079554/0.0477508) -- (2.175,0.014541/0.0477508);
        \draw [solid] (2.1625,0.014541/0.0477508) -- (2.1875,0.014541/0.0477508);
        \draw [solid] (2.1625,0.0079554/0.0477508) -- (2.1875,0.0079554/0.0477508);
        
        \draw [solid] (2.285,0.0055037/0.0477508) -- (2.285,0.0098456/0.0477508);
        \draw [solid] (2.2725,0.0098456/0.0477508) -- (2.2975,0.0098456/0.0477508);
        \draw [solid] (2.2725,0.0055037/0.0477508) -- (2.2975,0.0055037/0.0477508);
        
        \draw [solid] (2.5125,0.0252745/0.0477508) -- (2.5125,0.0482359/0.0477508);
        \draw [solid] (2.5,0.0482359/0.0477508) -- (2.525,0.0482359/0.0477508);
        \draw [solid] (2.5,0.0252745/0.0477508) -- (2.525,0.0252745/0.0477508);

        \draw [solid] (2.625,0.0149514/0.0477508) -- (2.625,0.0335147/0.0477508);
        \draw [solid] (2.6125,0.0335147/0.0477508) -- (2.6375,0.0335147/0.0477508);
        \draw [solid] (2.6125,0.0149514/0.0477508) -- (2.6375,0.0149514/0.0477508);
        
        \draw [solid] (2.7375,0.0103573/0.0477508) -- (2.7375,0.0312171/0.0477508);
        \draw [solid] (2.725,0.0312171/0.0477508) -- (2.75,0.0312171/0.0477508);
        \draw [solid] (2.725,0.0103573/0.0477508) -- (2.75,0.0103573/0.0477508);
    
        \draw [solid] (2.86375,0.0304874/0.0477508) -- (2.86375,0.0625559/0.0477508);
        \draw [solid] (2.85125,0.0625559/0.0477508) -- (2.875,0.0625559/0.0477508);
        \draw [solid] (2.85125,0.0304874/0.0477508) -- (2.875,0.0304874/0.0477508);    

        \draw [solid] (2.975,0.01846/0.0477508) -- (2.975,0.0344381/0.0477508);
        \draw [solid] (2.9625,0.0344381/0.0477508) -- (2.9875,0.0344381/0.0477508);
        \draw [solid] (2.9625,0.01846/0.0477508) -- (2.9875,0.01846/0.0477508);
        
        \draw [solid] (3.0875,0.0130148/0.0477508) -- (3.0875,0.0282039/0.0477508);
        \draw [solid] (3.1,0.0282039/0.0477508) -- (3.075,0.0282039/0.0477508);
        \draw [solid] (3.1,0.0130148/0.0477508) -- (3.075,0.0130148/0.0477508);
    
        \legend{\gls{FS}1, \gls{FS}2, \gls{FS}3}

    \end{axis}
\end{tikzpicture}
\caption{Scaled Winkler and Pinball of \gls{BNN} and \gls{QR} for S1-S3 and \gls{FS}1-\gls{FS}3 averaged over all estimated voltages. Min/max values are indicated by black bars.} \label{fig:Probabilistic}
\end{figure*}
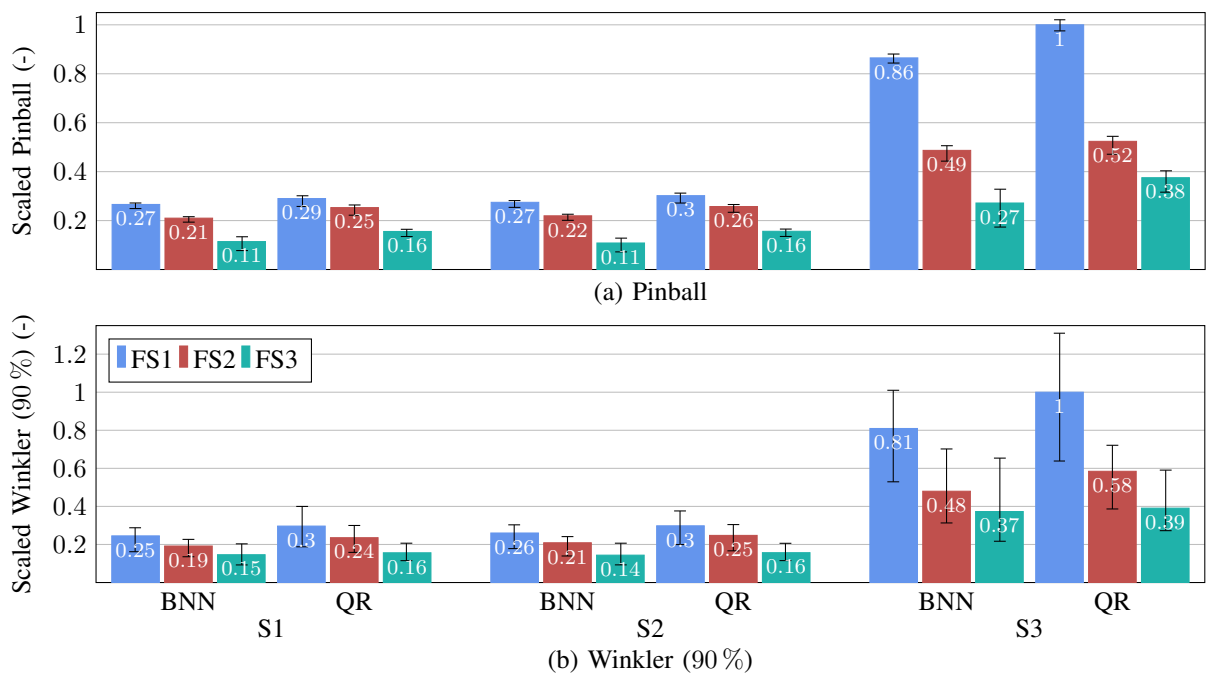 
\subsubsection{Point estimation performance}
The results in terms of \gls{RMSE} accuracy for the different flexibility scenarios and \glspl{FS} are shown in Fig. \ref{fig:RMSE}.
The scaled \gls{RMSE} values are plotted, with unity corresponding to an error of $0.0106$\,pu.

As assumed, point estimates improve with the level of observability (\gls{FS}1 to \gls{FS}3) in all flexibility scenarios. 
For S3 the decrease of \gls{RMSE} between \gls{FS}1 and \gls{FS}2 is comparatively large, as without real-time network measurements both the \gls{BNN} and \gls{QR} cannot accurately capture the frequent flexibility activations in the \gls{LV} network under consideration (see Fig. \ref{fig:FScomparison} (a)). 
Thus, the inclusion of primary substation information reduces \gls{RMSE} significantly by \SI{42}{\percent}.
As assumed from the qualitative evaluation, S1 and S2 show similar point estimation performance. 
However, a large relative \gls{RMSE} increase can be noticed for S3. 
The large presence of flexibility activations in the estimated \gls{LV} network increases load variability and uncertainty, complicating voltage estimation. 
Another finding is the superior performance of the \gls{BNN} compared to the \gls{QR} for all flexibility scenarios and \glspl{FS}. 
On average, the \gls{BNN} improves \gls{RMSE} by \SI{10.3}{\percent}. 

\subsubsection{Probabilistic estimation performance}
By considering Pinball and Winkler score, the probabilistic estimation performance is evaluated for reliability, sharpness and resolution. 
In accordance with the point estimation performance evaluation, the scores are shown in Fig. \ref{fig:Probabilistic} for S1-S3 and \gls{FS}1-\gls{FS}3. 
Values are scaled to a Pinball score of $0.00328$\ pu and a Winkler (\SI{90}{\percent}) score of $0.0478$ pu, respectively. 

A comparison between S1 and S2 shows that both the \gls{BNN} and \gls{QR} for all \glspl{FS} provide credible \glspl{PI} also in a scenario with frequent flexibility activation below adjacent secondary substations. 
Although for \gls{FS}2 both models, compared to S1, show a performance decrease in the range of \SI{5}{\percent}, it can be concluded that frequent flexibility activations below adjacent substations only slightly increase randomness, and thus barely deteriorate the correlation between primary substation measurements and bus voltages in the examined \gls{LV} network. 
By including secondary substation measurements (\gls{FS}3) this effect can be fully mitigated. 
In contrast, for frequent flexibility activations in the investigated \gls{LV} network (S3) both scores indicate a strong performance decrease for all \glspl{FS}.
However, for S3 the performance gain of the \gls{BNN} by including secondary substation measurements (\gls{FS}3) is comparatively large. 
While for S1 and S2 Pinball decreases by approximately 10 percentage points, a decrease by 22 percentage points is achieved in S3. 
A similar behavior can be derived from Winkler (\SI{90}{\percent}). 
As a result, for \gls{FS}3 the \gls{BNN} is able to keep both scores in a comparable performance range as for S1 and S2 with \gls{FS}1 and \gls{FS}2. 
It can be concluded that, especially under frequent flexibility activations in the same \gls{LV} network, incorporating secondary substation measurements adds great value to probabilistic \gls{LVSE}. 
Moreover, although Fig. \ref{fig:Probabilistic} shows a strong relative performance decrease for S3, from Fig. \ref{fig:FScomparison} it can be seen that already for \gls{FS}2 the \gls{BNN} provides credible \glspl{PI} that successfully capture sudden voltage drops.

Another finding is that the considered deterministic and probabilistic performance metrics (see Fig. \ref{fig:RMSE} and \ref{fig:Probabilistic}) show a similar behavior. 
Therefore, the \gls{BNN} outperforms the \gls{QR} benchmark across all scenarios, also in terms of probabilistic \gls{SE}. 
The average performance improvement for Pinball is \SI{17.2}{\percent} and for Winkler \SI{13.4}{\percent}.
In this context, it should be considered that the large size of the training dataset allowed to reduce epistemic uncertainty to a negligible part (see Fig. \ref{fig:FScomparison}). 
In cases of less training data or more frequently changing conditions, such as grid topology changes or newly added \gls{EV} fleets, it can be assumed that the \gls{BNN} by capturing both aleatoric and epistemic uncertainty has an even stronger advantage. 

\subsection{Evaluation of uncertainty behavior} \label{subsec:uncertainty_evaluation}
\begin{figure}[t]
\centering
    \input{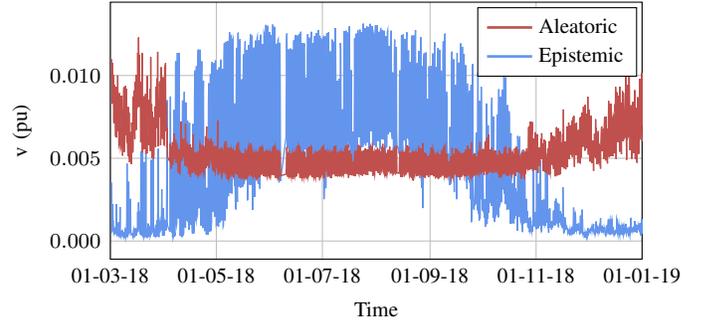}
    \caption{Epistemic and aleatoric uncertainty of S1 for a \gls{BNN} trained on January and February $2018$ with \gls{FS}2. 
    Uncertainties correspond to the \SI{90}{\percent} \gls{PI}.
    The few gaps are the result of short periods of missing \gls{SM} data.} \label{fig:uncertainty_evaluation}
\end{figure}

To shed light on the behavior of aleatoric and epistemic uncertainty under varying conditions, the \gls{BNN} was trained on January and February data, and used to estimate bus voltage $4$ for the remaining $10$ months of the year. 
Note that this is a showcase which intentionally does not consider model retraining. 
In Fig. \ref{fig:uncertainty_evaluation}, both uncertainty types are shown for S1 and \gls{FS}2. 
In accordance with the previous findings (see Fig. \ref{fig:FScomparison}), epistemic uncertainty is low for a recently trained model. 
However, large peaks can be noticed, which correlate with solar radiation and result from lack of \gls{PV} generation in the training data. 
Between April and October, additionally, an increasing trend of epistemic uncertainty is visible. This results from shifting away from the training data, and is induced by higher ambient temperatures. 
In practice, frequent retraining would keep epistemic uncertainty small during the entire year. 
From November on, epistemic uncertainty approaches zero again, as data moves towards the distribution of the training set. 
Due to little \gls{PV} generation, no peaks occur in December. 

Aleatoric uncertainty decreases between April and October. 
In fact, it exhibits a strong correlation with substation loadings, which are lower during this period because of lower heating demand. 
The fact that aleatoric uncertainty is higher for a recently trained model shows that, in contrast to epistemic uncertainty, frequent retraining cannot reduce it as it stems from randomness rather than a lack of knowledge. 

The following conclusions can be drawn. 
For a recently trained model, aleatoric uncertainty is highly dominant, justifying the application of models for \gls{LVSE} capable of capturing it. 
However, without the epistemic counterpart, uncertainty induced by sudden or ongoing changes is not quantified, which can be seen from the \gls{PV}-induced peaks and increasing trend, respectively, in Fig. \ref{fig:uncertainty_evaluation}. 
As such changes are present in \gls{LV} networks, a model which additionally considers epistemic uncertainty is beneficial for \gls{LVSE}, both for improved estimation accuracy, but also for better interpretability. 
While sudden peaks are an indicator for unknown events, an increasing trend of epistemic uncertainty indicates retraining need. 
Thus, by providing accurate quantification of both uncertainties, the \gls{BNN} is seen as a valuable approach for \gls{LVSE}.

\section{Conclusion} \label{sec:Conclusion}
In this work, uncertainty quantification in \gls{LVSE} is investigated for various flexibility usage scenarios. 
For that purpose, a \gls{BNN} capable of capturing both epistemic and aleatoric uncertainty is implemented and compared to a \gls{QR} benchmark.
Estimation and uncertainty quantification performance are systematically evaluated from a qualitative and quantitative perspective, and for multiple scenarios of input information availability and flexibility utilization, using real data. 
Results show that flexibility activations below adjacent secondary substations only have minor impact on \gls{LVSE}, while activations in the same network decrease performance significantly. 
However, it is also shown that including secondary substation measurements allows retaining an acceptable degree of performance. 
By dynamically extending the \gls{PI} during flexibility activation periods, the \gls{BNN} is able to capture flexibility-induced voltage drops, enabling a more reliable \gls{LVSE}.
Moreover, it shows superior performance compared to the \gls{QR} benchmark for all considered cases. 
By considering and differentiating between epistemic and aleatoric uncertainty, it improves interpretability, as it provides insights in occurrence of unknown events or retraining need. 
Future work will be directed towards applying the \gls{BNN} for uncertainty quantification in \gls{LV} state forecasting.
\section*{Acknowledgment}
This work is partially supported by the INTERPRETER project, which has received funding from the European Union’s Horizon 2020 research and innovation programme under grant agreement No 864360, and ERA-NET project HONOR, funded under grant agreement No 646039 and No 7759750 (RegSys 2018).

\ifCLASSOPTIONcaptionsoff
  \newpage
\fi
\bibliographystyle{ieeetr}

\end{document}